\documentclass{article}
\usepackage[utf8]{inputenc}

\usepackage[linktocpage=true]{hyperref}

\usepackage{xcolor}
\hypersetup{
	colorlinks,
	linkcolor={black},
	citecolor={black},
	urlcolor={black}
}
\usepackage{amsmath}
\usepackage[english]{babel}
\usepackage{graphicx}
\usepackage{makeidx}
\usepackage[usenames,dvipsnames]{pstricks}
\usepackage{setspace}
\usepackage{xcolor}
\usepackage{sectsty}
\usepackage{textcomp}
\usepackage{amssymb}
\usepackage{floatflt}
\usepackage{epsfig}
\usepackage{tocstyle}

\usetocstyle{standard}

\newcommand{\vn}{\vskip .5cm\noindent}
\newcommand{\nn}{\noindent}
\newcommand{\bs}{\begin{subequations}}
	\newcommand{\es}{\end{subequations}}

\newcommand{\ra}{\rightarrow}

\newcommand{\lla}{\langle}
\newcommand{\rra}{\rangle}

\newcommand{\sig}{\sigma}
\newcommand{\lam}{\lambda}

\newcommand{\alp}{\alpha}
\newcommand{\be}{\beta}

\makeatletter

\usepackage{amsfonts}
\usepackage{pdfpages}

\usepackage{setspace}

\usepackage{babel}
\usepackage{bookman,fancyhdr}
\usepackage{exscale}

\usepackage[linktocpage=true]{hyperref}

\makeatother

\usepackage{tocloft}

\begin{document}
	
	
	
	
	
	
	
	
	\title{\centerline{\textbf{\Large Universality in Quantum Measurements}}}
	
	\vskip 4cm
	
	\author{{\Large{\textbf{Avijit Lahiri}}}\\{\centerline{Department of Physics, Vidyasagar Metropolitan College, Kolkata 700006}}\footnote{email: avijit.lahiri.al@gmail.com; blog (TacitKnowledge): tacit-views.blogspot.com}}
	
	\vskip 3cm

	\date{}
	
	\maketitle

	\begin{abstract}
		
		\vn We briefly review a number of major features of the approach to quantum measurement theory based on environment-induced decoherence of the measuring apparatus, and summarize our observations in the form of a couple of general principles that, unlike the wave function collapse hypothesis, emerge as ones consistent with the unitary Schr\"odinger evolution of wave functions. We conclude with a few observations of a philosophical nature, to the effect that that quantum theory does not purport to describe reality but constitutes an {\it interpretation} of our phenomenal reality within a context --- one where the Planck scale is not crossed. Beyond the Planck scale, a radically new interpretation of reality is likely to emerge.

		
		
		
	\end{abstract}

\vn {\bf Keywords:} quantum measurement, decoherence, relative entropy, entanglement, quantum correlations, classical correlation, mixed states of a classical object, nearest separable state 

\vn {\bf PACS:} 03.65.Ta, 03.65.Yz, 03.67.Bg, 03.67.Mn


\newpage

\section{Introduction}\label{intro-sec}

\vn Decoherence has gained a place of central relevance in the theoretical formulation of quantum mechanics, notably so because it promises to replace the wave function collapse postulate with principles compatible with the rest of the basic assumptions of the theory. Indeed, environmental decoherence generates hope for  a solution (of sorts) to the so-called ‘measurement problem’. In this brief note we state a general principle that implies a universality in the process of environment-induced effacement of the system-apparatus entanglement (see below) in the measurement process. Decoherence theory was pioneered by Zeh, Zurek, Paz, and other contemporary workers (see ~\cite{Schlosshauer} for a useful bibliography, covering the period from the origin of the theory up to a time when it gained wide acceptance).

\vn A measurement in the present context involves a quantum mechanical system (S), a measuring apparatus (A), {\it and} the {\it environment} (E), where E operates indirectly but crucially through its effect on A, the latter being a quantum mechanical system {\it in the limit of classicality}. While making no claim to originality, we make explicit (based on~\cite{Lahiri}) the role of E in the measurement process {\it through its effect on A}, which is commonly not analyzed in detail in the literature. The decoherence effect of E is essentially one of {\it global entanglement sharing}.     

\section{Relative entropy and entanglement}\label{entang-sec}

\vn To start with, we note that the concept of quantum {\it relative entropy} can be invoked to introduce a measure of separation (though not in the strict sense of a metric) between two states (represented by density operators, say, $\hat\rho,\hat\sig$) of a system,

\begin{align}
S(\hat\rho||\hat\sigma)={\rm Tr}(\hat\rho\ln\hat\rho)-{\rm Tr}(\hat\rho\ln\hat\sigma), ~~\label{relative-eq1}
\end{align}

\nn since this tells us how distinguishable $\hat\rho$ is from $\hat\sigma$. This same quantity {\it also} gives us a measure of the degree of quantum correlation, or {\it entanglement}, of a bipartite system in a state $\hat\rho$ (\cite{Vedral-Plenio}, \cite{Vedral}): if, according to the above measure of separation, $\hat{\bar\rho}$ be the {\it closest separable} state to the state $\hat\rho$ of a bipartite system made up of subsystems, say, A,B, then the entanglement measure (referred to subsystems A,B) in $\hat\rho$ is, simply, $S(\hat\rho||\hat{\bar\rho})$. Incidentally, the concept of entanglement measure is not a uniquely defined one, and other measures may be introduced, all having a number of essential common features that any reasonable definition of quantum correlations is to possess --- for instance, analogous definitions are possible on the basis of the Renyi relative entropy and the Bures Metric (\cite{Vershynina}, \cite{Vedral-Plenio}). In the present paper, though, our considerations will be based on the quantum relative entropy. The idea underlying the definition of quantum correlations as a distance to the closest separable state can be extended to include multipartite systems as well.

\vn This same concept of the relative entropy can be made use of in quantifying the {\it classical} correlations between subsystems of a composite system (\cite{Vedral-Plenio}, \cite{Henderson-Vedral}). For instance, considering again the composite system (say, C) made up of subsystems A,B, the classical correlation (referred to subsystems A,B) in the state $\hat\rho$ of C can be defined as the relative entropy between the closest separable state $\hat{\bar\rho}$ and the completely uncorrelated  direct product state ${\hat{\bar\rho}}_{\rm A}\otimes{\hat{\bar\rho}}_{\rm B}$ involving the {\it reduced} states ${\hat{\bar\rho}}_{\rm A},~{\hat{\bar\rho}}_{\rm B}$. 

\vn Generally speaking, the fact that the quantum correlations in a state (such as $\hat\rho$ above) of a composite system (C in the present instance) signify something over above the classical correlations, is observed experimentally as a violation of the Bell-type inequalities, though the relation between entanglement and the Bell inequalities is a nuanced one (see, e.g.,~\cite{Werner}), especially in the context of multipartite states. 

\section{Decoherence and the measurement process}\label{process-sec}

\vn Entanglement is generated and shared dynamically between systems by means of interactions between them in the course of their unitary Schr\"odinger evolution. However, considered in the context of any subsystem that is itself a composite system, the reduced state of the former shows up non-unitary decoherence and relaxation effects. This is what happens in a quantum measurement where one has a measured system (say, S), a measuring apparatus (A), and the environment (E) in mutual interaction. Among these, A and E are systems characterized by an enormously large number of degrees of freedom each. In particular, A can be considered to be {\it a quantum mechanical system close to the classical limit}. Bohr had, in early days (especially in addressing the Einstein-Podolsky-Rosen critique), underlined the crucial role (in the measurement process) played by the classical nature of the measuring apparatus. Von Neumann built upon Bohr's point of view by way of looking at the apparatus as a quantum mechanical system (\cite{Schlosshauer}) having classical features, and advanced his seminal measurement model where the system (S) got entangled with the apparatus (A), with specific states of S getting associated with corresponding specific states of A. However, the quasi-classical nature of A was not explicitly given a place of relevance in the theory till the advent of a point of view where environment-induced decoherence assumed a position of fundamental significance.


\vn Specifically, the interaction between A and E has the effect that states of A are {\it mixed} ones, characterized by specific values of a {\it pointer variable}, where the latter has to correspond in some appropriate way to the observable meant to be measured by A. We will later briefly look at the setting up of pointer states of a classical object in the process of environment-induced decoherence. For now, we consider for simplicity a 2D system S with an observable $\hat s$ having eigenvalues $s_1,s_2$ and corresponding eigenstates $|s_1\rra,|s_2\rra$, where the {\it measurement interaction} between S and A generates an entangled state in which $|s_1\rra,|s_2\rra$ get associated with pointer states $\hat{ M}_1, \hat{ M}_2$ of A (\cite{Schlosshauer}; see~\cite{Qureshi} for a concrete example; the Pointer states have nearly classical properties --- thus, each of two conjugate variables have nearly well-defined values). In order that this association be realized by means of the measurement interaction, one has to have a measuring apparatus A that is {\it appropriate} for the measurement of the observable $\hat s$ under consideration. For instance, some other apparatus (say, A$'$) may have pointer states that do not get associated with $|s_1\rra,|s_2\rra$ by an appropriate S-A interaction. The quasi-classical nature of A ensures that ${\hat{ M}}_1, {\hat{ M}}_2$ are mixtures made up of enormous numbers of pure states.

\section{Decoherence to the nearest separable state}\label{nearest-sec}

\vn A simplified model of the measurement process was considered in~\cite{Lahiri} where the measurement interaction generates a system-apparatus entangled state of the form 

\begin{align}
\hat\rho=|\alp_1|^2|s_1\rra\lla s_1|\otimes{\hat{ M}_1}+|\alp_2|^2|s_2\rra\lla s_2|\otimes{\hat{ M}_2}
+\hat{\cal{I}} +{\hat{\cal{I}}}^\dagger.~~\label{mixed-eq1}
\end{align} 

\nn In this expression, $\alp_1,\alp_2$ are complex amplitudes defining the pure state of S in which $\hat s$ is sought to be measured, and ${\hat {\cal I}},{\hat{ {\cal I}}}^\dagger$ are {\it interference terms} accounting for the system-apparatus entanglement (an explicit representation is given in~\cite{Lahiri} in the simplified measurement model considered there). These interference terms introduce a correlation between $\hat{ M}_1,\hat { M}_2$ that gets obliterated by environmental decoherence since quantum correlations between states of a classical object do not survive the decoherence process (see below). As the correlation between $\hat{ M}_1,\hat { M}_2$ is destroyed, the system-apparatus entanglement also gives way to the separable reduced S-A state

\begin{align}
\hat{\bar\rho}=|\alp_1|^2|s_1\rra\lla s_1|\otimes{\hat{ M}_1}+|\alp_2|^2|s_2\rra\lla s_2|\otimes{\hat{ M}_2}
,~~\label{mixed-eq2}
\end{align}

\nn where only the {\it classical} correlation between S and A remains --- the probabilities $|\alp_1|^2,|\alp_2|^2$ now pertain to {\it pre-existing} values $s_1,s_2$ associated with the pointer states ${\hat M}_1,{\hat M}_2$ of the measuring apparatus. This represents the `post-measurement pre-selection' S-A state where the result of measurement has not been read out by noting the pointer state that the apparatus is in --- such recording constitutes a classical observation on A. If, on the other hand, the pointer state is observed and noted to be, say, ${\hat M}_i~(i=1,2)$ (this occurs with probability $|\alp_i|^2$) then the S-A state reduces to $|s_i\rra\lla s_i|\otimes\hat{M_i}$ --- the `post-measurement post-selection' state. In other words, the measurement process can be schematically described as a succession: measurement interaction $\ra$ S-A entangled state $\ra$ environment-induced decoherence of entangled states of the apparatus $\ra$ the post-measurement pre-selection state $\ra$ classical observation on A and recording of the measurement result $\ra$ post-measurement post-selection state (see~\cite{Venugopalan}, where a concrete model is worked out in detail). In this scheme the measurement interaction and the decoherence process can be distinguished only notionally --- both occur simultaneously where, in particular, the process of decoherence occurs {\it instantaneously} in virtue of the quasi-classical nature of the measuring apparatus. However, this distinction is significant in that the state~\eqref{mixed-eq2} can be shown to be the {\it nearest separable state} to the entangled state~\eqref{mixed-eq1}.

\vn We note that the process of environmental decoherence amounts to wiping off the off-diagonal terms in the density matrix (written in terms of appropriate basis states of S and A) represented by $\hat{\cal{I}}, {\hat{\cal{I}}}^\dagger$ in~\eqref{mixed-eq1}. Though apparently an irreversible process, it is actually a unitary one, realized by means of an effectively random interaction between A and E (see sec.~\ref{two-sec} below) --- one that involves a global entanglement sharing by degrees of freedom pertaining to E. 

\vn This measurement scheme (described in the context of the simple model outlined in~\cite{Lahiri}), in which the Von Neumann approach (see~\cite{Schlosshauer}) is carried to its logical conclusion by the incorporation of the decoherence effect of the environment on the apparatus, promises to be a complete description of the quantum measurement process, assuming that the following three questions are addressed concretely: (a) how the measurement interaction actually generates the S-A entanglement in specific instances of measurement, (b) why the apparatus always exists in classical pointer states, which are mixtures of enormous numbers of microscopic pure states, and (c) why the environment-induced decoherence process is, to all intents and purposes, an instantaneous one. We briefly address below the second of the above three issues in an attempt to clarify the commonly encountered statement to the effect that a macroscopic classical system `cannot exist in a superposed state'. A general principle pertaining to the first question is not available (however, particular models exist), while the third issue will also be briefly addressed later in the paper. 

\section{States of a classical object}\label{classical-sec}

\vn In reality, the quanum mechanical description of states of a classical system is a non-trivial matter. For instance, referring to the Schr\"{o}dinger cat, one often states that a superposition of $|\rm{live~cat}\rra$ and $|\rm{dead~cat}\rra$ quickly degenerates by decoherence. On examining how the two `states' referred to are related, one notes that they differ in terms of a number of `vital parameters' such as body temperature, heart rate, and parameters relating to the functioning of various bodily organs. These define a number of `macroscopic' variables that contrast with an enormous number of microscopic variables relating to the states of molecules in the body of the cat. More generally, the state of a macroscopic classical object such as a solid body is described in terms of a number of macroscopic variables (such as the shape and size of the solid body) along with an enormously large number microscopic ones, where the former --- relatively few in number --- are in the nature of {\it collective} variables in contrast to the latter. The state space of such a body can be looked upon as a direct product of the space accommodating the collective states and one pertaining to the microscopic states, where the dimension of the latter space is enormously large compared to the former.

\vn In other words we can, in a manner of speaking, describe the states of a classical object (such as a measuring apparatus) as those belonging to a direct product, say, ${\cal C}\otimes{\cal M}$, where $\cal C$ is the space of the collective variables and $\cal M$ corresponds to the microscopic ones. To make things simple, let us assume that states in $\cal M$ correspond to discrete values of a single pointer variable, two such states being, say,  $|\lam_1\rra$ and $|\lam_2\rra$. A pointer value such as $\lam_1$ or $\lam_2$ does not, however, describe the state of the system (the apparatus A in the present instance) as a whole and one has to specify, in addition, the states of the microscopic variables, say, $s_1,s_2,\cdots, s_N$, where $N$ is an enormously large number. Thus, a `superposition' of pointer states $|\lam_1\rra$ and $|\lam_2\rra$, can only mean a state in ${\cal C}\otimes{\cal M}$ of the form

\begin{align}
|\xi\rra=\sum_{\{u\}}\alp_{\{u\}}|\lam_1;\{u\}\rra+\sum_{\{v\}}\beta_{\{v\}}|\lam_2;\{v\}\rra,~~ \label{mixed-eq3}
\end{align}

\nn where $\{u\},\{v\}$ represents $N$-tuples of values of variables $s_1,s_2,\cdots, s_N$, $\alp_{\{u\}}, \beta_{\{v\}}$ are associated amplitudes, and the sums are over all possible values of the $N$-tuples.

\vn However, because of the interaction with the environment that is, to all intents and purposes, a random one, and of the large dimension of $\cal M$, phase correlations between states with distinct values of the microscopic variables are destroyed in no time and one is left with the disentangled mixed state

\begin{align}
\hat M=\sum_{\{u\}}|\alp_{\{u\}}|^2|\lam_1;\{u\}\rra\lla\lam_1;\{u\}|+\sum_{\{v\}}|\be_{\{v\}}|^2|\lam_2;\{v\}\rra\lla\lam_2;\{v\}|,~~ \label{mixed-eq4}
\end{align}

\nn In other words, a superposition of pure states with specified values of the pointer variable $\lam$ degenerates into an admixture of mixed states with those same values of $\lam$, the latter being in the nature of conserved (or nearly conserved) quantities. Put differently, a pointer state with some specified value of the pointer variable is a mixed state involving an enormous number of microscopic states --- states ${\hat M}_1, {\hat M}_2$ appearing in~\eqref{mixed-eq1}, \eqref{mixed-eq2}  are precisely such states of the measuring apparatus A. It may be mentioned here that the pointer variables of a classical object A are determined by the interaction between it and the environment E, being `selected' by the latter, depending on approximate conservation principles satisfied by the interaction.

\vn The transition from~\eqref{mixed-eq3} to~\eqref{mixed-eq4} is made possible by essentially the same process as in~\eqref{mixed-eq1} to~\eqref{mixed-eq2}, involving an effecively random interaction indicative of a global sharing of entanglement. 

\vn A comparison with the process of measurement on S by means of A considered above indicates that, in a manner of speaking, the process of environmental decoherence giving rise to a mixed state of a classical object is analogous to the one in which the post-measurement pre-selection S-A state~\eqref{mixed-eq2} is realized in a measurement process. Put differently, in the case of a classical object, the set of microscopic variables in $\cal M$ acts as an `apparatus', and the quantum correlations between the `pointer system' in $\cal C$ and the `apparatus' get destroyed by the effect of the environment, consequent to which the environment itself `reads off' the value of the pointer variable, generating the post-selection state --- one like ${\hat M}_1$ or ${\hat M}_2$ considered above in the case of the measurement apparatus A. 

\vn The emergence of classicality in macroscopic objects by means of environmental decoherence is in the nature of a limiting one since exact classicality is in contrast to quantum principles, as seen by results relating to contextuality and non-locality (see, for instance,~\cite{Mermin}). In this context, one notes that exact classicality is limited by the principle of {\it macrorealism} too where the time evolution of a quantum mechanical system is seen to violate the Leggett-Garg inequality (refer to~\cite{Leggett}) that constrains the time evolution of a system under the assumption of classicality.  

\section{The measurement process: universal principles}\label{two-sec}

\vn At this point we go back to the measurement process and indicate two major features of the process of environmental decoherence in which the classically correlated S-A state given by~\eqref{mixed-eq2} is generated.

\vn {\it First}, it occurs in an infinitesimally small time interval, i.e., can be assumed to be {\it instantaneous} as compared to all other relevant time scales involving quantum mechanical systems. For instance, it is orders of magnitude smaller than the time scales by which individual quantum mechanical systems evolve under mutual interactions and {\it also} the time scales over which microscopic systems ordinarily suffer environment-induced decoherence effects. This is essentially due to the enormously large number of the microscopic degrees of freedom of the apparatus and to the fact that decoherence does not involve transitions between the microscopic states by the impact of environmental degrees of freedom (which is what happens in the relaxation process) but requires just a dephasing of the natural oscillation frequencies of these states.

\vn {\it Secondly}, the measure of quantum correlation between the system and the apparatus that is required to be removed in the process of environmental decoherence (transforming the state~\eqref{mixed-eq1} to \eqref{mixed-eq2}) is {\it infinitesimally small} (see~\cite{Lahiri}) when compared with the {\it classical} correlation that remains in~\eqref{mixed-eq2} as compared with the completely uncorrelated direct product state resulting from the read-out process in which some particular measured value ($s_1$ or $s_2$ in the present instance) is selected. A necessary condition for this is that the mixed pointer states ${\hat M}_1$ and ${\hat M}_2$ of the apparatus are to be  distinguishable to an adequate degree, i.e., the corresponding sets of weights (such as $\{|\alp_{\{u\}}|^2\}$ and $\{|\be_{\{v\}}|^2\}$ appearing in~\eqref{mixed-eq4}) are sufficiently distinct (as measured in terms of their relative entropy) --- in the simple model considered in~\cite{Lahiri}, ${\hat M}_1, {\hat M}_2$ reside in orthogonal subspaces of the space of microscopic states of A. Such distinguishability, indeed, is a necessary condition for the measurement to produce a meaningful result. 

\vn In summary of these observations, we suggest that the quantum measurement process is governed by the following simple but general principles: {\it the S-A entanglement between the measured system and the measuring apparatus created by the measurement interaction is erased by the environmental decoherence of the apparatus, which is a quasi-classical object, where the measure of entanglement so erased (expressed in terms of separation based on relative entropy: the S-A system goes over to the nearest separable state) is infinitesimally small as compared to the classical correlation that remains in the S-A system till some particular measurement result is read off. In addition, the removal of entanglement occurs in an infinitesimally small time interval}. The principles are general in that these do not refer to any particular features of the measured system or the measuring apparatus (assuming that there is an appropriate match between the two). As is apparent, the two principles are related to each other.

\vn The same principles apply, in particular, when the measured system (S) is itself in an entangled state, such as the singlet state of two spin-$\frac{1}{2}$ particles, the two being at a large separation from one another. Denoting the spin-up and spin-down states of the $i$th particle  by $|\pm\rra_i~(i=1,2)$, and assuming that only the second particle is measured with the help of the apparatus, one can construct a putative S-A entangled state (generated by the measurement interaction) along the lines of~\cite{Lahiri}, in which case the post-measurement pre-selection state would be of the form

\begin{align}
\hat\rho=\frac{1}{2}|+\rra_1 ~ _1\lla +| \otimes |-\rra_2 ~ _2\lla -|\otimes{\hat{ M}_1}+\frac{1}{2}|-\rra_1~_1\lla - |\otimes|+\rra_2~_2\lla +|\otimes{\hat{ M}_2},~~\label{mixed-eq5}
\end{align}

\nn where environmental decoherence once again eliminates the interference terms, thereby erasing the quantum correlation between S and A (notably, a similar result would be obtained if the apparatus were to measure the first spin alone). As before, this quantum correlation would be infinitesimally small compared to the classical correlation that would be erased in case of a read-out of the measurement result. 

\vn The generality of the principles enunciated above assumes that the environment-induced decoherence of the apparatus states results from the most general form of the apparatus-environment interaction, namely one based on a {\it random matrix} interaction Hamiltonian (see~\cite{Gorin} for background; also, as a concrete application, see~\cite{Wang}). Such an interaction was assumed in~\cite{Lahiri}, where the large dimensions of the apparatus and the environment ensured the right course of decoherence even for a relatively weak strength of interaction. It is the same generality of the above principles that ensures that all results emanating from the wave function collapse hypothesis follow from the decoherence approach as well. In particular, the features of non-locality and contextuality in a classical formulation of these results in terms of hidden variables follows as a necessary consequence.                    

\vn The above scheme of quantum measurements, based on the idea of environment-induced decoherence (for a closely related approach, see~\cite{Vedral}), is a closed one in which there is no  need for the wave function collapse hypothesis since all the relevant interactions, including the interactions with the environment, generate unitary Schr\"{o}dinger evolution --- it is only the infinitesimally small time scale associated with the removal of quantum correlations that generates the impression of wave function collapse.

\vn However, the scheme just falls short of being a truly closed one in that a realistic estimate (of general validity) of the time scale of the environmental decoherence process for quantum mechanical systems close to the classical limit is yet to emerge. Indeed, numerous estimates including ones taking into consideration the role of gravitation and quantum field fluctuations at ultra-short length scales indicate that the time scale is likely to be be as small as the {\it Planck time} for classical objects (see, e.g.,~\cite{Petruzziello}, \cite{Arzano}, \cite{Diosi}; the very assumption of the existence of fluctuations associated with a fundamental length is seen to lead to an effective decoherence mechanism characterized by the Planck scale). If so, radically new principles will come up, placing quantum mechanics itself in a new perspective --- the `measurement problem' will then be seen to be a mere symptom of an incompleteness at a deeper level that gets exposed by means of the decoherence approach. The decoherence approach itself will then be updated in novel ways that physics will discover beyond the Planck scale.

\vn There exists a large literature devoted to whether and to what extent the decoherence theory solves the quantum measurement problem or, more generally, the problems perceived to attach to quantum theory itself (see, in this context,~\cite{Adler}, \cite{Schlosshauer1}). This, of course, depends to some extent on the point of view one adopts as to what it is precisely that constitutes the `problem' and what exactly is acceptable as a `solution' to it. The present paper does not presume to contribute to that literature but proposes a framework whereby the measurement postulate can be reconciled with unitary time evolution (while, at the same time, pointing to a couple of general principles characterizing the measurement process), while the catch to it lies in the uncertainties associated with considerations relating to time scales short enough to compare with the Planck scale. The process of decoherence by means of field fluctuations (whereby entanglement is shared globally) points to the frontier that, perhaps, is to be encountered in this endeavor to make quantum theory `problem-free'. However, as indicated below (sec.~\ref{reality-sec}) theories, in crossing frontiers, become problem-free only at the cost of being insidiously contaminated with stranger problems.

\section{Quantum mechanics as a theory of reality}\label{reality-sec}

\vn Einstein, along with his collaborators, spoke of an {\it incompleteness} of quantum theory in their E-P-R critique. In the present article, we have looked at a {\it lack of coherence} in the theory where the measurement postulate is seen to be at odds with the Schr\"odinger evolution of a quantum mechanical system. The decoherence approach seems to promise to bring back coherence, but is itself in need of a clear formulation in respect of the decoherence time --- the latter, being likely to be of the order of the Planck time, may need new concepts in order to be precisely defined. In other words, the `lack of coherence' may indeed, at the end, turn out to signify an `incompleteness'. However, compared to the E-P-R critique, that incompleteness now appears in a new light --- it is to be set right not necessarily by trying to restore `objectivity' to quantum theory (say, by means of hidden variables), but by the formulation of a fresh interpretation of reality where, at the same time, quantum field theory and gravitation may acquire a new significance.

\vn Science does not bear the responsibility of {\it describing} reality as it is, or of unearthing `laws' inherent in nature. While reality exists in itself beyond our mental efforts at comprehension, we build up fractional {\it interpretations} of that reality in making up our {\it phenomenal} world(see~\cite{Lahiri1}). Quantum theory, like any other theory in science, formulates explanations of events and their correlations as observed in that phenomenal world of ours. All theories are provisional and defeasible since they require radical revisions as the {\it context} of our observations changes. In the case of quantum theory and gravitation, that context is set by the Planck scale.

\pagestyle{plain}

\end{document}